\documentclass[a4paper,11pt]{article}

\pdfoutput=1
\usepackage{amsmath}
\usepackage{graphicx}

\usepackage[english]{babel}
\usepackage[utf8]{inputenc}
\usepackage{pdflscape}
\usepackage{enumerate}
\usepackage{amsbsy}
\usepackage{amsmath} 
\usepackage{graphics}
\usepackage{mathrsfs}
\usepackage{wrapfig}
\usepackage{mathtools}

\usepackage{amsfonts}
\usepackage{pstricks}
\usepackage{color}
\usepackage{setspace}

\usepackage{accents}
\usepackage{tensor}

\usepackage{jcappub_ver} 

\newcommand{\bea}{\begin{eqnarray}} \newcommand{\eea}{\end{eqnarray}}
\newcommand{\el}{\nonumber \\}
\newcommand{\re}[1]{(\ref{#1})}

\newcommand{\pat}{\partial}

\renewcommand{\sec}[1]{section \ref{#1}}

\newcommand{\tab}[1]{table \ref{#1}}

\renewcommand{\a}{\alpha}
\renewcommand{\b}{\beta}
\renewcommand{\c}{\gamma}
\renewcommand{\d}{\delta}

\newcommand{\m}{\mu}
\newcommand{\n}{\nu}

\newcommand{\ha}{\frac{1}{2}}

\newcommand{\rmd}{\mathrm{d}}

\newcommand{\ie}{i.e.\ }
\newcommand{\eg}{e.g.\ }

\newcommand{\bx}{\boldsymbol{x}}
\newcommand{\bk}{\boldsymbol{k}}

\newcommand{\sg}{\sqrt{-g}}

\title{Inflation with the Chern--Simons term in the Palatini formulation}

\author{Ali Hassan}
\author{and Syksy R\"{a}s\"{a}nen}

\affiliation{University of Helsinki, Department of Physics and Helsinki Institute of Physics \\ P.O. Box 64, FIN-00014 University of Helsinki, Finland}

\emailAdd{ali.hassan@helsinki.fi}
\emailAdd{syksy.rasanen@iki.fi}

\abstract{
We consider the Chern--Simons term coupled to the inflaton in the Palatini formulation of general relativity. In contrast to the metric formulation, here the Chern--Simons term affects also the background evolution. We approximately solve for the connection, insert it back into the action, and reduce the order of the equations to obtain an effective theory in the gradient approximation. We consider three cases: when the connection is unconstrained, and when non-metricity or torsion is put to zero. In the first two cases, the inflaton kinetic term is modified with a term proportional to the square of the potential. For polynomial potentials dominated by the highest power of the field, the Chern--Simons term solves the problem that higher order corrections spoil the flatness of the potential. For Higgs inflation, the tensor-to-scalar ratio can be as large as the current observational bound, and the non-minimal coupling to the Ricci scalar can be as small as in the metric case. The Palatini contribution cures the known instability of the tensor modes due to the Chern--Simons term in the metric formulation.
}

\begin{document}

\begin{flushleft}
	\hfill		 HIP-2024-18/TH \\
\end{flushleft}
 
\setcounter{tocdepth}{3}

\setcounter{secnumdepth}{3}

\maketitle

\section{Introduction} \label{sec:intro}

Inflation is the most successful scenario for the early universe \cite{Starobinsky:1979ty, Starobinsky:1980te, Kazanas:1980tx, Guth:1981, Sato:1981, Mukhanov:1981xt, Linde:1981mu, Albrecht:1982wi, Hawking:1981fz, Chibisov:1982nx, Hawking:1982cz, Guth:1982ec, Starobinsky:1982ee, Sasaki:1986hm, Mukhanov:1988jd}, with predictions in excellent agreement with observations \cite{Akrami:2018odb, BICEP:2021xfz}. The simplest candidate to drive inflation is a single scalar field, called the inflaton. Such a field is expected to couple non-minimally to the curvature, as shown by general effective field theory arguments \cite{Weinberg:2008hq, Solomon:2017nlh} and explicit loop calculations \cite{Callan:1970ze}. Coupling to the Ricci scalar is a key ingredient in Higgs inflation \cite{Bezrukov:2007, Bezrukov:2013, Bezrukov:2015, Rubio:2018ogq}, and more complicated couplings have also been considered \cite{Capozziello:1999uwa, Capozziello:1999xt, Daniel:2007kk, Sushkov:2009hk, Germani:2010gm, Germani:2010ux, Kobayashi:2010cm, Kamada:2010qe, Kobayashi:2011nu, Germani:2011mx, Tsujikawa:2012mk, Kamada:2012se, Kamada:2013bia, Germani:2014hqa, Yang:2015pga, Kunimitsu:2015faa, DiVita:2015bha, Escriva:2016cwl, Fumagalli:2017cdo, Sato:2017qau, Fu:2019ttf, Granda:2019wyi, Granda:2019wip, Sato:2020ghj, Fumagalli:2020ody, Nezhad:2023dys}.

Particularly interesting are topological terms such as the Gauss--Bonnet term and the Chern--Simons term (also called the Pontryagin term), both quadratic in the Riemann tensor \cite{Lue:1998mq, Jackiw:2003pm, Alexander:2009tp}. They can be motivated by string theory and loop quantum gravity, and for the Chern--Simons term also by a gravitational anomaly in the lepton current in the Standard Model \cite{Alexander:2004us, Alexander:2004xd, Lyth:2005jf, Alexander:2006lty, Alexander:2009tp, Alexander:2016hxk}. As boundary terms, they do not alone contribute to the classical equations of motion, but become dynamical when coupled to a scalar field. Because the Chern--Simons term is parity-odd, it could possibly explain \cite{Creque-Sarbinowski:2023wmb} the recently claimed detection of parity violation in the four-point function of large-scale structure \cite{Hou:2022wfj, Philcox:2022hkh} (see also \cite{Krolewski:2024paz}). Unlike the Gauss--Bonnet term, the Chern--Simons term coupled to a scalar field is likely unstable, and must be treated in terms of an effective theory \cite{Delsate:2014hba, Crisostomi:2017ugk}.

In the metric formulation of general relativity, the Chern--Simons term depends only on the Weyl tensor, not on the Ricci tensor. Even when coupled to a scalar field, its contribution to the background evolution and the evolution of first order scalar perturbations during inflation vanishes. To leading order it thus modifies only gravitational waves, leading to polarisation-dependent propagation \ie birefringence, and an instability for one polarisation \cite{Lue:1998mq, Choi:1999zy, Alexander:2004us, Alexander:2004xd, Alexander:2004wk, Lyth:2005jf, Alexander:2006lty, Alexander:2009tp, Dyda:2012rj, Alexander:2016hxk, Bartolo:2017szm, Bartolo:2018elp, Sulantay:2022sag, Creque-Sarbinowski:2023wmb}.

In contrast, in the Palatini formulation of general relativity, where the connection is an independent degree of freedom, the Chern--Simons term gives a non-vanishing contribution also to the background and scalar perturbations.\footnote{In the Palatini formulation, there are also the Holst and the Nieh--Yan terms that are linear in the curvature and quadratic in the connection, and which contribute when coupled to the inflaton \cite{Langvik:2020nrs, Shaposhnikov:2020gts}.} In the Palatini formulation, all components of the connection can be left to be determined from the equations of motion, or some constraints can be imposed a priori \cite{Einstein:1925, Hehl:1976, Hehl:1978, Papapetrou:1978, Hehl:1981, ferraris1982, Percacci:1991, Percacci:2009, Percacci:2020bzf}. Most studied examples include putting the non-metricity to zero (called the Einstein--Cartan formulation), or putting the torsion to zero. Sometimes the latter is taken to be part of the definition of the Palatini formulation, while the case without a priori constraints is referred to as the metric-affine formulation. The Chern--Simons term in the Palatini formulation has been studied in \cite{Hehl:1990ir, Babourova:1996id, BottaCantcheff:2008pii, Alexander:2009tp, Sulantay:2022sag, Boudet:2022nub, Bombacigno:2022naf}.

We for the first time study coupling of the Chern--Simons term to the inflaton in the Palatini formulation. In \sec{sec:dof} we set up the effective theory of the Chern--Simons term by solving for the connection, inserting back into the action and reducing the order of the equations of motion. We consider the case when the connection is unconstrained, and the cases when either non-metricity or torsion is put to zero. In \sec{sec:infl} we apply the effective action to slow-roll inflation for some example models, calculate the corrections to the scalar and tensor power spectra, and show that the metric case Chern--Simons tensor instability is cured. In \sec{sec:conc} we summarise our results and comment on open problems.

\section{Chern--Simons term coupled to a scalar field} \label{sec:dof}

\subsection{Geometrical quantities} \label{sec:geom}

\subsubsection{Non-metricity, torsion, and curvature}

In the Palatini formulation the metric $g_{\a\b}$ and the connection $\Gamma^\c_{\a\b}$ are independent variables. We decompose the connection, defined with the covariant derivative as $\nabla_\b A^\a\equiv\pat_\b A^\a + \Gamma^\a_{\b\c} A^\c$, $\nabla_\b A_\a\equiv\pat_\b A_\a - \Gamma^\c_{\b\a} A_\c$, as
\bea \label{Gamma}
  \Gamma^\c_{\a\b} &=& \mathring\Gamma^\c_{\a\b} + L^\c{}_{\a\b} = \mathring\Gamma^\c_{\a\b} + J^\c{}_{\a\b} + K^\c{}_{\a\b} \ ,
\eea
where $\mathring\Gamma_{\a\b}^\c$ is the Levi--Civita connection of the metric $g_{\a\b}$, and $L^\c{}_{\a\b}$ is the distortion tensor. We denote quantities defined with the Levi--Civita connection by $\mathring{}$. In the second equality we have decomposed the distortion $L^\c{}_{\a\b}$ into the disformation $J_{\a\b\c}$ and the contortion $K_{\a\b\c}$, defined as
\bea \label{JK}
  J_{\a\b\c} &\equiv& \frac{1}{2} \left(Q_{\a\b\c}  - Q_{\c\a\b} - Q_{\b\a\c} \right) \ , \qquad K_{\a\b\c} \equiv \ha (T_{\a\b\c} + T_{\c\a\b} + T_{\b\a\c} ) \ ,
\eea
where $Q_{\a\b\c}$ and $T_{\a\b\c}$ are the non-metricity and the torsion, respectively, defined as
 \bea \label{TQ}
  \qquad Q_{\c\a\b} \equiv \nabla_\c g_{\a\b} = - 2 L_{(\a|\c|\b)} \ , \qquad T^\c{}_{\a\b} &\equiv& 2 \Gamma^{\c}_{[\a\b]} = 2 L^\c{}_{[\a\b]} \ .
\eea
We have $Q_{\c\a\b}=Q_{\c(\a\b)}$, $J_{\a\b\c}=J_{\a(\b\c)}$, and $K^\c{}_\a{}^\b=K^{[\c}{}_\a{}^{\b]}$. The two non-metricity vectors are defined as
\bea \label{Qvec}
  Q^\c \equiv g_{\a\b} Q^{\c\a\b} \ , \qquad \hat Q^\b \equiv g_{\a\c} Q^{\a\b\c} \ ,
\eea
and the torsion vector and torsion axial vector are defined as, respectively,
\bea \label{Tvec}
  T^\b \equiv g_{\a\c} T^{\a\b\c} \ , \qquad \hat T^\a \equiv \frac{1}{6} \varepsilon^{\a\b\c\d} T_{\b\c\d} \ ,
\eea
where $\varepsilon_{\a\b\c\d}$ is the Levi--Civita tensor.\footnote{We include pseudotensors under the label tensor; likewise for vectors and pseudovectors, and scalars and pseudoscalars.}

The Riemann tensor can be decomposed into the Levi--Civita part and the distortion part as
\bea \label{Riemann}
  \!\!\!\!\!\!\! R^{\a}{}_{\b\c\d} &\equiv& \pat_\c \Gamma^{\a}_{\d\b}-\pat_\d\Gamma^{\a}_{\c\b} + \Gamma^{\a}_{\c\mu} \Gamma^{\mu}_{\d\b} - \Gamma^{\a}_{\d\mu}\Gamma^{\mu}_{\c\b}= \mathring R^{\a}{}_{\b\c\d} + 2 \mathring \nabla_{[\c} L^\a{}_{\d]\b} + 2 L^\a{}_{[\c|\mu|} L^\mu{}_{\d]\b} \ .
\eea
The Ricci tensor is $R_{\a\b}\equiv R^{\c}{}_{\a\c\b}$, and the Ricci scalar is $R\equiv g^{\a\b}R_{\a\b}$.

\subsubsection{Chern--Simons term}

The Chern--Simons term is \cite{Jackiw:2003pm}
\bea \label{CS}
\frac12 
  {^*}R^{\a}{}_{\b\c\d} R^{\b}{}_{\a}{}^{\c\d} 
  &\equiv& 
  \frac14 \varepsilon_{\c\d}{}^{\mu\nu} R^{\a}{}_{\b\mu\nu} R^{\b}{}_{\a}{}^{\c\d} = \mathring\nabla_\a K^\a \ ,
\eea
where ${^*}R^{\a}{}_{\b\c\d}\equiv\ha\varepsilon_{\c\d}{}^{\mu\nu} R^{\a}{}_{\b\mu\nu}$ is the dual Riemann tensor\footnote{In \cite{Boudet:2022nub} an alternative definition of the Chern--Simons term was used, which includes the homothetic curvature tensor $R^{\c}{}_{\c\a\b}$ so that the total term is invariant under the projective transformation \cite{Hehl:1978}, while remaining a total derivative. In the metric formulation, it reduces to the same Chern--Simons term, but in the Palatini formulation it leads to different physics.}, and $\mathring\nabla_\a K^\a\equiv\frac{1}{\sg}\pat_\a(\sg K^\a)$. The Chern--Simons term is antisymmetric under parity transformations and (separately) time reversal, so it is a pseudoscalar. In the second equality we have made it transparent that the Chern--Simons term is a total derivative by introducing the Pontryagin current
\bea \label{K}
  K^\a &\equiv& 
 \varepsilon^{\a \b \c \d}
    \left(
    \Gamma^\mu_{\b \nu}
    \partial_\c 
    \Gamma^\nu_{\d \mu}
    +
    \frac23
    \Gamma^\mu_{\b \nu}
    \Gamma^\nu_{\c \lambda}
    \Gamma^\lambda_{\d \mu}
    \right) \el
  &=&
 \mathring{K}^\a
+
 \varepsilon^{\a \b \c \d}
    \left(
    \tensor{L}{^\mu_{\b \nu}}
    \tensor{\mathring{R}}{^\nu _\mu _\c _\d}
    +
    \tensor{L}{^\mu_{\b \nu}}
    \mathring\nabla_\c 
    \tensor{L}{^\nu_{\d \mu}}
      +
    \frac23
    \tensor{L}{^\mu_{\b \nu}}
    \tensor{L}{^\nu_{\c \lambda}}
    \tensor{L}{^\lambda_{\d \mu}}
    \right)
 \ .
\eea
On the second line we have applied the decomposition \re{Riemann} to separate $K^\a$ into $\mathring K^\a$ that is present also in the metric case and the contribution that is new to the Palatini case. The Palatini term is a tensor, but $\mathring K^\a$ is not, as indicated by the fact that it can be put to zero at any point by choosing coordinates so that $\mathring\Gamma^\c_{\a\b}=0$ there. Hence $K^\a$ is not a vector, although as \re{K} shows its divergence, defined in terms of the coefficients $K^\a$ as if they were the components of a vector, is a scalar.

\subsection{Action and equations of motion} \label{sec:act}

\subsubsection{Action} \label{sec:action}

As the Chern--Simons term is a total derivative, it does not contribute to the classical equations of motion by itself. We consider a scalar field coupled to the Chern--Simons term, which makes it dynamical, with the action (we use units where the reduced Planck mass is unity)
\bea \label{action}
  S &=& \int\rmd^4 x \sqrt{-g} \left[ \ha R - \ha K(\varphi) g^{\a\b} \pat_\a \varphi \pat_\b \varphi - V(\varphi) + \frac{1}{4} P(\varphi) \varepsilon^{\c\d\mu\nu} R^{\a}{}_{\b\mu\nu} R^{\b}{}_{\a\c\d} \right] \el
  &=& \int\rmd^4 x \sqrt{-g} \left( \ha R - \ha K X - V - K^\a\pat_\a P \right) \ ,
\eea
where we have included the non-minimal kinetic function $K(\varphi)$ (not to be confused with $K^\a$) and the non-minimal Chern--Simons or Pontryagin coupling $P(\varphi)$, and we denote $X_{\a\b}\equiv\pat_\a\varphi\pat_\b\varphi$, $X\equiv X^\a{}_\a$. On the second line we have dropped a boundary term (we drop boundary terms from the action also in what follows). The resulting Lagrangian is not a scalar, but as it is a scalar up to boundary terms, its variation is a tensor. Including a non-minimal coupling to the Ricci scalar of the form $F(\varphi) R$ would, via a conformal transformation of the metric, be equivalent to the redefinition $K\to K/F$, $V\to V/F^2$, so we can neglect it without loss of generality. The Chern--Simons term is invariant under the conformal transformation. If $\varphi$ sits in the minimum of its potential, the contribution of the Chern--Simons term vanishes, and there are no observational constraints on it.

Were we to demand the action to be invariant under parity transformations, we would have to choose $\varphi$ to be a pseudoscalar and $P(\varphi)$ to be odd. If this is not done, the theory breaks parity and time reversal invariance in the scalar-gravity sector. These symmetries are broken in the Standard Model, and there seems to be little reason to ask them to be respected here. (The Chern--Simons term does not break CPT symmetry.) Hence we do not restrict $\varphi$ to be a pseudoscalar. It could be the Higgs field of the Standard Model, a possibility we take up in \sec{sec:higgs}.

In the action \re{action}, the Riemann tensor \re{Riemann} depends only on the connection, not on the metric. We will vary the action with respect to the connection, approximately solve its equation of motion, and insert the solution back into the action to obtain an action that depends only on $g_{\a\b}$ and $\varphi$. We consider three physically distinct cases: when no constraints are imposed on the connection, and when either non-metricity or torsion is put to zero a priori.

\subsubsection{Unconstrained case}

Let us first solve for the connection in the case when there are no a priori constraints. Varying the action \re{action} with respect to the distortion tensor $L_{\a\b\c}$ gives the equation
\bea \label{LEOM}
  L_{\a\b\c} + L_{\c\a\b} - g_{\b\c} L_{\a\mu}{}^\mu - g_{\a\c} L^\mu{}_{\mu\b} = 2 P_{\a\b\c} \ ,
\eea
where we have defined
\bea \label{P}
  P_{\a\b\c} &\equiv& 2 {^*}R_{\a\b\c}{}^\lambda \pat_\lambda P = ( \mathring R_{\a\b\mu\nu} + 2 \mathring \nabla_{\mu} L_{\a\nu\b} + 2 L_{\a\mu\lambda} L^\lambda{}_{\nu\b} ) \varepsilon^{\mu\nu}{}_\c{}^{\lambda} \pat_\lambda P \ ,
\eea
where we have applied the decomposition \re{Riemann}. The fact that the Chern--Simons term is quadratic in the derivatives of the connection leads to the covariant derivative of $L_{\a\b\c}$ in \re{P}, so \re{LEOM} is not an algebraic but a differential equation for the connection. In the metric formulation, the Chern--Simons term coupled to a scalar field leads to third order equations for the metric, and the system likely suffers from the Ostrogradsky instability \cite{Delsate:2014hba, Crisostomi:2017ugk}. In the Palatini case, the equations are second order for $\varphi$, first order for $L_{\a\b\c}$, and algebraic for $g_{\a\b}$.\footnote{Note that the Chern--Simons term in \re{action} does not depend on the metric because the dependence in $\sqrt{-g}$ and $\varepsilon^{\a\b\c\d}$ cancels, so it does not contribute to the metric equation of motion.} However, this does not imply that the system is free of ghosts. No stability analysis of the Chern--Simons term coupled to a scalar field in the Palatini formulation has been done, but there is no obvious reason to expect it to be stable.

Indications of instability can be seen by treating the action as an effective theory and the Chern--Simons contribution as a perturbative correction. As it is proportional to $\pat_\a P=\pat_\a \varphi P'$ (where prime denotes derivative with respect to $\varphi$) we can either consider $P'$ to be small, or allow $P'$ to be large but take $\pat_\a \varphi$ to be small, as is true in the super-Hubble regime during slow-roll inflation. In either case, we can solve \re{LEOM} perturbatively, taking $P_{\a\b\c}$ to be small.

At first sight, the structure of \re{LEOM} seems to present a problem for such an expansion. The highest (in this case first) order derivative is multiplied with the small parameter, so the order of the differential equation changes as the parameter goes to zero, indicating that the expansion is singular, and the solutions are not perturbative in the expansion parameter \cite{Simon:1990ic, Simon:1990jn, Parker:1993dk}. At the lowest order, \re{LEOM} is algebraic, but at second order in $\pat_\a P$ it turns into a first order equation. However, the singular structure is avoided due to the antisymmetry of the Levi--Civita tensor. If $\pat_\a P$ is timelike (as is the case during inflation in the super-Hubble regime), then \re{LEOM} does not contain time derivatives of $L_{\a\b\c}$, only spatial derivatives. Then the equation remains algebraic as regards time evolution (with a non-trivial constraint structure related to the spatial derivatives), and can be solved order by order in $\pat_\a P$.

We consider the solution to leading order in $\pat_\a P$; during slow-roll inflation in the super-Hubble regime, higher order terms are suppressed. We then have $P_{\a\b\c}\simeq\mathring P_{\a\b\c}\equiv2 {^*}\mathring R_{\a\b\c}{}^\lambda \pat_\lambda P$ (note that $\mathring P_{\a\b\c}=\mathring P_{[\a\b]\c}$, and that $\mathring P_{\a\b\c}$ is traceless), and the solution of \re{LEOM} is
\bea \label{Lsol}
  L_{\a\b\c} &=& \mathring P_{\a\b\c} - \mathring P_{\c\a\b} + \mathring P_{\b\c\a} \ .
\eea
Non-metricity is zero, $Q_{\a\b\c}=0$, there is only torsion, $T_{\a\b\c}=2\mathring P_{\b\c\a}$. At this leading order, where $P_{\a\b\c}$ does not depend on the connection, \re{LEOM} also allows the homogenous solution $L_{\a\b\c}=g_{\a\c} A_\b$, where $A_\a$ is an arbitrary vector, corresponding to the projective mode \cite{Hehl:1978}. The Chern--Simons term is not invariant under the projective symmetry $\Gamma^\a_{\b\c} \to \Gamma^\a_{\b\c} + \delta^\a{}_\c A_\b$, except trivially at leading order (when it does not contain the connection). The full solution of \re{Lsol} thus does not contain an arbitrary projective mode, and the term in the solution with the same structure is determined by the source terms at second order. As we consider the solution only to first order, we do not include such a term.

At higher orders, the non-linearity of \re{LEOM} will generate higher powers of $\pat_\a\varphi$ and $\mathring R_{\a\b\c\d}$ in the solution for $L_{\a\b\c}$. Due to the antisymmetry of $\varepsilon_{\a\b\c\d}$ in \re{P}, higher than first order time derivatives of $\varphi$ will not be generated in $L_{\a\b\c}$ at any order. But when inserting $L_{\a\b\c}$ back into the action \re{action}, the term $\mathring\nabla_\c\tensor{L}{^\nu_{\d \mu}}$ in $K^\a$ in \re{K} will generally give second order time derivatives of $\varphi$. They can lead to unstable new degrees of freedom, as can the terms higher order in the Riemann tensor. We neglect such possible extra degrees of freedom as beyond the validity of our effective theory. If we went to higher orders, then from the effective theory point of view it could be sensible to also include higher powers of the Chern--Simons term in the action; they have been studied in the metric case \cite{Crisostomi:2017ugk}.

Taking the solution \re{Lsol}, inserting it back into the action \re{action}, and expanding the Einstein--Hilbert and the Chern--Simons term to quadratic order in $P$, we obtain
\bea \label{actionphi}
  S &=& \int\rmd^4 x \sqrt{-g} \left[ \ha \mathring R - \ha K X - V + \frac{1}{4} P \varepsilon_{\c\d}{}^{\mu\nu} \mathring R^{\a}{}_{\b\mu\nu} \mathring R^{\b}{}_{\a}{}^{\c\d} + P'{}^2 X_{\a\b} ( \a_1 \mathring R^{\a\m\n\lambda} \mathring R^\b{}_{\m\n\lambda} \right. \el
  && \left. + \a_2 g^{\a\b} \mathring R^{\m\n\lambda\sigma} \mathring R_{\m\n\lambda\sigma}
  + \a_3 \mathring R^{\a\m} \mathring R^\b{}_{\m} 
  + \a_4 g^{\a\b} \mathring R^{\mu\n} \mathring R_{\m\n} 
  + \a_5 \mathring R^{\a\m\b\n} \mathring R_{\m\n} ) \right] \ ,
\eea
where the coefficients $\a_n$ are given in \tab{tab:abc}. The term proportional to $P$ is the same as in the metric formulation, whereas the terms proportional to $P'{}^2$ appear only in the Palatini formulation. Like the Palatini term, the metric theory term depends only on $P'$ (not $P$), but it is more convenient to write it in terms of $P$ and the tensor $\mathring R_{\a\b\c\d}$, rather than $P'$ and the non-tensorial quantity $\mathring K^\a$. The metric theory term violates parity if $P$ is parity-even, but the Palatini term does not. At third order in $P'$ there would be another parity-odd term, at fourth order another parity-even term, and so on.

\begin{table*}[t!]
  \begin{center}
  \begin{tabular}{|c|c|c|c|}
  \hline
  Coefficient & Unconstrained & Zero non-metricity & Zero torsion \\
  \hline
  $\a_1$ & $-8$ & $-8$ & $-5$ \\
  $\a_2$ & $3$ & 3 & 2 \\
  $\a_3$ & $4$ & 4 & 2 \\
  $\a_4$ & $-4$ & $-4$ & $-2$ \\
  $\a_5$ & $8$ & 8 & 4 \\
  \hline
  \end{tabular}
  \end{center}
  \caption{Coefficients $\a_n$ in the action \re{actionphi} for the cases when there are no a priori constraints on the connection, and when non-metricity or torsion is assumed to vanish.} 
  \label{tab:abc}
\end{table*}

\subsubsection{Zero non-metricity}

The case when the non-metricity is put to zero a priori is often called Einstein--Cartan theory. Taking into account that now $L^\c{}_\a{}^\b=L^{[\c}{}_\a{}^{\b]}$, the equation for the distortion is \re{LEOM} antisymmetrised in $\a$ and $\b$. The solution to leading order has been considered in \cite{BottaCantcheff:2008pii, Alexander:2009tp}. As non-metricity happens to vanish in the unconstrained case, we here get the same solution \re{Lsol} to leading order. In general, imposing a constraint a priori leads to a different theory (see \eg \cite{BeltranJimenez:2019acz, Annala:2022gtl, Nezhad:2023dys}) so the solutions are not expected to agree at higher orders.

\subsubsection{Zero torsion}

Putting the torsion to zero a priori, \ie assuming that the connection is symmetric, is often taken as part of the definition of the Palatini formulation, with the unconstrained case called metric-affine theory. The equation for the distortion is now \re{LEOM} symmetrised in $\a$ and $\c$. The leading order solution is
\bea \label{Qsol}
  L_{\a\b\c} = 2 \mathring P_{\a(\b\c)} \ ,
\eea
and the non-metricity is equal to the distortion, $Q_{\a\b\c}=2 \mathring P_{\a(\b\c)}$. Inserting the solution back into the action \re{action}, we get the action \re{actionphi} with the coefficients $\a_n$ listed in \tab{tab:abc}. The differences in the distortion lead to a different action for the inflaton once we transfer the physics of the distortion to the scalar sector, which we will now do.

\subsubsection{Order reduction}

The action \re{actionphi} contains terms quadratic in the Riemann tensor and is not of the Horndeski nor DHOST form \cite{Horndeski:1974wa, Langlois:2018dxi, Kobayashi:2019hrl}, so it involves higher than second derivatives and is unstable. As noted, we treat even the starting action \re{action} as an effective field theory, and take the instabilities to fall outside its domain of validity. We thus reduce the order of the differential equations to remove the spurious instabilities \cite{Bel:1985zz, Simon:1990ic, Simon:1990jn, Simon:1991bm, Parker:1993dk, Weinberg:2008hq, Delsate:2014hba, Solomon:2017nlh}. We decompose the Levi--Civita Riemann tensors in the action  \re{actionphi} into Weyl and Ricci pieces,
\bea
  \mathring R_{\a\b\c\d} = \mathring C_{\a\b\c\d} + ( g_{\a[\c} \mathring R_{\d]\b} - g_{\b[\c} \mathring R_{\d]\a} ) - \frac{1}{3} g_{\a[\c}g_{\d]\b} \mathring R \ ,
\eea
to obtain the action
\begin{align} \label{actionWeyl}
  S =& \int\rmd^4 x \sqrt{-g} \bigg\{ \ha \mathring R - \ha K X - V + \frac{1}{4} P \varepsilon_{\c\d}{}^{\mu\nu} \mathring C^{\a}{}_{\b\mu\nu} \mathring C^{\b}{}_{\a}{}^{\c\d}
\nonumber \\ & \nonumber
+ P'{}^2 X_{\a\b} \Big[ - \frac16 \left( \a_1 + 2 \a_2 + \a_5 \right) g^{\a\b} \mathring R^2 
  + \frac13 \left( \a_1 + 2 \a_5 \right) \mathring R^{\a\b} \mathring R 
  + \left( \a_3 - \a_5 \right) \mathring R^\a{}_\mu \mathring R^{\b\mu} 
\\ & \nonumber
  + \frac12 \left( \a_1 + 4 \a_2 + 2 \a_4 + \a_5 \right) g^{\a\b} \mathring R^{\mu\nu} \mathring R_{\mu\nu}
  + \left( 2 \a_1 + \a_5 \right) \mathring C^{\a\mu\b\nu} \mathring R_{\mu\nu}
\\ & 
  + \a_1 \mathring C^{\a\mu\nu\lambda} \mathring C^\b{}_{\mu\nu\lambda} 
  + \a_2 g^{\a\b} \mathring C^{\mu\nu\lambda\sigma} \mathring C_{\mu\nu\lambda\sigma} \Big] \bigg\} \ .
\end{align}
There is no Ricci contribution to the last term on the first line, the only one that is present in the metric formulation, due to the antisymmetry of the Levi--Civita tensor. The non-minimally coupled Chern--Simons tensor thus does not affect the equations of motion in conformally flat spacetimes in the metric formulation. In particular, this is true for the Friedmann--Lema\^{\i}tre--Robertson--Walker (FLRW) model. In contrast, the Palatini pieces, which come from solving the connection, involve the Ricci tensor. We use the equation of motion of the metric to write the Ricci tensor in terms of the energy-momentum tensor and get rid of the associated higher order derivatives that are beyond the range of validity of the effective field theory. Varying the action \re{actionWeyl} with respect to the metric and working to zeroth order in $P'$, we get the usual metric formulation Einstein equation
\bea \label{Einstein}
  \mathring R_{\a\b} - \ha g_{\a\b} \mathring R = T_{\a\b} \ ,
\eea
with the energy-momentum tensor
\bea \label{T}
  T_{\a\b} = K X_{\a\b} - g_{\a\b} \left( \ha K X + V \right) \ .
\eea
From \re{Einstein} we have $\mathring R_{\a\b} = T_{\a\b} - \ha g_{\a\b} T$, where $T\equiv T^\a{}_\a=-K X-4V$. Inserting this into the action \re{actionWeyl} and using \re{T}, we obtain
\bea \label{actionfinal}
   S &=& \int\rmd^4 x \sqrt{-g} \left[ \ha \mathring R - V - \ha K X 
   + \left( \frac23 \a_1 + \frac83 \a_2 + \a_3 + 4 \a_4 +\a_5 \right)
   P'{}^2 V^2 X \right. \el
   && 
   + \left( \frac43 \a_1 + \frac43 \a_2 + 2 \a_3 + 2 \a_4 + \a_5 \right) P'{}^2 K V X^2 
   + \left( \frac23 \a_1 + \frac53 \a_2 + \a_3 + \a_4 \right) P'{}^2 K^2 X^3 \el
   && \left. + P'{}^2 X_{\a\b} ( \a_1 \mathring C^{\a\mu\nu\lambda} \mathring C^\b{}_{\mu\nu\lambda} + \a_2 g^{\a\b} \mathring C^{\mu\nu\lambda\sigma} \mathring C_{\mu\nu\lambda\sigma} ) + \frac{1}{4} P \varepsilon_{\c\d}{}^{\mu\nu} \mathring C^{\a}{}_{\b\mu\nu} \mathring C^{\b}{}_{\a}{}^{\c\d} \right] \ .
\eea
We have now moved the effect of the Chern--Simons term to the scalar sector and the Weyl tensor. The scalar part is of the Horndeski form and thus stable (for appropriate choices of $K$ and $V$), but the Weyl terms are neither Horndeski nor DHOST, and would contain ghosts were they not treated as an effective field theory \cite{Horndeski:1974wa, Langlois:2018dxi, Kobayashi:2019hrl}.

In the unconstrained and the zero non-metricity case, the action \re{actionfinal} is (inserting the coefficients $\a_n$ from \tab{tab:abc})
\bea \label{actionfinalgen}
   \!\!\!\!\!\!\!\! S &=& \int\rmd^4 x \sqrt{-g} 
   \left[
   \ha \mathring R 
   - V 
   - \ha \left( K + \frac{8}{3} P'{}^2 V^2 \right) X 
   + \frac43 P'{}^2 K V X^2 
   - \frac13 P'{}^2 K^2 X^3 
   \right. \el
     && \left. 
   + P'{}^2 X_{\mu\nu} 
   ( - 8 \mathring C^{\mu\a\b\c} \mathring C^\nu{}_{\a\b\c} 
   + 3 g^{\m\n} C^{\a\b\c\d} \mathring C_{\a\b\c\d} 
   ) 
   + \frac{1}{4} P \varepsilon_{\c\d}{}^{\mu\nu} \mathring C^{\a}{}_{\b\mu\nu} \mathring C^{\b}{}_{\a}{}^{\c\d} 
   \right] \ .
\eea
The leading order Chern--Simons contribution is the modification of the kinetic term by the additive term proportional to $P'{}^2 V^2$. In contrast, in the case when the torsion is taken to be zero a priori, the kinetic term is not modified, and the action \re{actionfinal} reduces to
\bea
   \!\!\!\!\!\!\!\! S &=& \int\rmd^4 x \sqrt{-g} 
   \left[
   \ha \mathring R 
   - V 
   - \ha K X 
   \right. 
   \el
   && 
   \left. 
   + P'{}^2 X_{\mu\nu} 
   ( 
   - 5 \mathring C^{\mu\a\b\c} \mathring C^\nu{}_{\a\b\c} 
   + 2 g^{\m\n} C^{\a\b\c\d} \mathring C_{\a\b\c\d} 
   ) 
   + \frac{1}{4} P \varepsilon_{\c\d}{}^{\mu\nu} \mathring C^{\a}{}_{\b\mu\nu} \mathring C^{\b}{}_{\a}{}^{\c\d} 
   \right] \ .
\eea
Let us now apply the action \re{actionfinal} to slow-roll inflation.

\section{Inflation} \label{sec:infl}

\subsection{Background and scalar perturbations} \label{sec:scalar}

\subsubsection{Slow-roll}

In inflation, the equations of motion are split into the FLRW background and perturbations. In the metric formulation, where the Chern--Simons contribution reduces to the last term of \re{actionfinal}, the background evolution is obviously unchanged, as the Weyl tensor is zero, and the evolution of linear scalar perturbations is also unaffected. In the Palatini case, it remains true that the Weyl terms do not affect the background nor the scalar perturbations at leading order in the gradient approximation. The Weyl terms are quadratic in second spatial derivatives of the scalar perturbation, and do not contain time derivatives. They can therefore affect the evolution only at large wavenumber, and are small in the super-Hubble limit. So the evolution of the background and the leading order scalar perturbations is fully captured by the scalar field part of the action. This is not the case at higher orders, so there could be non-standard non-Gaussian signatures not captured by the scalar field kinetic terms and the potential. There are also (as in the metric case) modifications to tensor perturbations, which we discuss in \sec{sec:tensor}.

If we consider the theory in the gradient approximation (\ie assume that both space and time derivatives of the field are small), the validity of the terms higher order in $X$ in the action \re{actionfinal} is questionable. In any case, in slow-roll $K |X|\ll V$, so \re{actionfinal} reduces to the minimally coupled scalar field action with a kinetic term linear in $X$,
\bea \label{actioncan}
   S &=& \int\rmd^4 x \sqrt{-g} \left( \ha \mathring R - \ha \tilde K X - V \right) \ ,
\eea
where in the unconstrained and the zero non-metricity case we have
\bea \label{tildeK}
  \tilde K &\equiv& K + \frac{8}{3} P'{}^2 V^2 \ .
\eea
In the zero torsion case we just have $\tilde K=K$. We concentrate on the unconstrained and the zero non-metricity case. Note that if the solution \re{Lsol} for the connection is considered perturbative in the gradient $\pat_\a\varphi$ and not in $P'$, the term proportional to $P'^2$ in \re{tildeK} does not have to be a small correction to $K$. The result is reminiscent of New Higgs Inflation and other models where the coupling of the kinetic term to the Einstein tensor (and in the Palatini case, to other contractions of the Riemann tensor) generates a contribution proportional to $V$ in $\tilde K$ \cite{Germani:2010gm, DiVita:2015bha, Escriva:2016cwl, Fumagalli:2017cdo, Gialamas:2020vto, Nezhad:2023dys}. As the Chern--Simons term is quadratic in the Riemann tensor, we get $P'{}^2 V^2$ instead, and the modification is generated by the order reduction procedure, there is no direct coupling to the kinetic term to begin with. Let us illustrate the effect of this term with some examples of the kinetic function $K$, potential $V$, and non-minimal Chern--Simons coupling $P$. 

\subsubsection{Minimal coupling to Ricci scalar}

The Chern--Simons term is dimension 4, and the coupling function makes it higher order. We consider only terms even in $\varphi$ (as is the case if the field comes from a doublet, like the Standard Model Higgs). Let us first consider the lowest order term possible, $P=\ha P_0\varphi^2$, where $P_0$ is a constant, and take $K=1$, $V=V_0\varphi^n$, where $V_0$ and $n$ are constants. Then
\bea \label{tildeK1}
  \tilde K &=& 1 + \frac{8}{3}P_0^2 V_0^2 \varphi^{2n+2} \ .
\eea
For large field values, the canonical scalar field is
\bea
  \chi &=& \int\rmd\varphi \sqrt{\tilde K} \simeq \sqrt{\frac{2}{3}} \frac{2}{n+2} P_0 V_0 \varphi^{n+2} \ ,
\eea
so $\varphi\propto\chi^{\frac{1}{n+2}}$, and the potential is
\bea
  V &\propto& \chi^{\frac{n}{n+2}} \ .
\eea
Such a potential gives the slow-roll parameters
\bea \label{SR}
  \epsilon &\equiv& \ha \left( \frac{V'}{V} \right)^2 = \frac{n^2}{2 (n+2)^2} \chi^{-2} \el
  \eta &\equiv& \frac{V''}{V} = - \frac{2 n}{(n+2)^2} \chi^{-2} \ ,
\eea
so the spectral index is
\bea \label{ns1a}
  n_s &=& 1 - 6 \epsilon + 2 \eta = 1 - \frac{n ( 3 n + 4 )}{( n + 2 )^2} \chi^{-2} \ .
\eea
The number of e-folds is
\bea \label{N1}
  N &=& \int_{\chi_\textrm{end}}^\chi \frac{\rmd\chi}{\sqrt{2\epsilon}} = \frac{n+2}{2 n} \chi^2 \ ,
\eea
where we have assumed $\chi\gg\chi_\textrm{end}$; the subscript refers to the end of inflation. Inserting $\chi$ in terms of $N$ from \re{N1} into \re{ns1a}, we obtain
\bea \label{ns1b}
  n_s &=& 1 - \frac{3 n + 4}{2 (n + 2)} \frac{1}{N} \ .
\eea

Quadratic potential, $n=2$, gives $V\propto\chi^{1/2}$. For $n=4$, as in the case of the Standard Model Higgs, we get $V\propto\chi^{2/3}$. If we include terms in the potential up to $\varphi^6$ (motivated by the fact that the Chern--Simons contribution is also dimension 6) and assume that it dominates, we have $n=6$ and $V\propto\chi^{3/4}$. As $n$ grows, the potential approaches the linear case $V\propto\chi$, for which $n_s=0.970$ (for $N=50$); the dependence on $n$ is weak. The quadratic case gives $n_s=0.975$, while $n=4$ and $n=6$ both give $n_s=0.973$. A more careful accounting of $N$, considering violation of slow-roll towards the end of inflation and the duration of preheating, can shift the values. Also, if the second term in \re{tildeK1} does not dominate, the potential interpolates between $\chi^{\frac{n}{n+2}}$ and $\chi^n$, moving the spectral index down.

If we instead take both the potential and the Chern--Simons contribution to have the same dimension, we get $V\propto\varphi^n$, $P\propto\varphi^{n-4}$. This gives $V\propto\chi^{\frac{n}{2n-4}}$, which again interpolates between $V\propto\chi$ (for $n=4$) and $V\propto\chi^{1/2}$ (for $n\to\infty$), corresponding to $n_s=0.975$ and $0.970$ (again taking $N=50$), respectively. The non-minimal coupling to the Chern--Simons term thus solves the problem that higher order contributions to the inflaton potential would spoil the flatness of the potential. A similar mechanism operates if there is a non-minimal coupling between the inflaton and a gauge field in the Palatini formulation of gauge field theory \cite{Rasanen:2022ijc}.

The above values of $n_s$ are consistent with observations, as is well known for monomial potentials $V\propto\chi^\a$ with $\a<1$ \cite{Akrami:2018odb, BICEP:2021xfz}. So the non-minimal Chern--Simons term can make a quadratic potential, or any other potential dominated by a single power of the field, consistent with observations as far as the spectral index (and its running) is concerned. However, the model also has to respect the upper bound on the amplitude of tensor perturbations. In usual canonical minimally coupled single-field slow-roll inflation the tensor-to-scalar ratio is $r=16\epsilon$, which is above the observational upper bound $r<0.036$ for $\a>0.45$ \cite{BICEP:2021xfz}. If more than one power of $\varphi$ contributes to the potential when the observable perturbations are generated, a more careful calculation would be needed, and $n_s$ and $r$ could change. In our case, the Chern--Simons term also modifies the evolution of gravitational waves, although as we will see, the change to $r$ is small in the range of validity of our approximation.

\subsubsection{Non-minimal coupling to Ricci scalar} \label{sec:higgs}

Let us then consider the case when the original action \re{action} includes the non-minimal coupling $\ha F(\varphi) R$, as in Higgs inflation \cite{Bezrukov:2007, Bauer:2008}. As noted in \sec{sec:action}, it can be shifted to the scalar sector with a conformal transformation of the metric that leaves the Chern--Simons term invariant but gives $K\to K/F$, $V\to V/F^2$. We thus end up with the action \re{actioncan} with the potential $V/F^2$ and the kinetic coupling
\bea
  \tilde K &=& \frac{K}{F} + \frac83 P'{}^2 \frac{V^2}{F^4} \ .
\eea
As in the original Higgs inflation proposal, we take $K=1$, the Standard Model potential $V=\frac{1}{4}\lambda\varphi^4$, and the dimension 4 non-minimal coupling $F=1+\xi\varphi^2$. Taking $P=\ha P_0\varphi^2$, where $P_0$ is a constant, and assuming that the largest powers of $\varphi$ dominate, $V/F^2$ is a constant, so we have $\tilde K\propto\varphi^2$. This leads to $\chi\propto\varphi^2$, so the asymptotically flat potential is $V/F^2\simeq\frac{\lambda}{4\xi^2}( 1 - \frac{\kappa}{\chi} )$, where $\kappa\equiv\lambda P_0/(\sqrt{6}\xi^3)$. Without the coupling $P$, Higgs inflation leads to an asymptotically exponentially flat potential, which in the Palatini case has the large coefficient $\sqrt{\xi}$ in the exponent, and the correspondingly small tensor-to-scalar ratio $r=\frac{2}{\xi N^2}$ \cite{Bezrukov:2007, Bauer:2008}. The normalisation of the CMB perturbations gives (for $N=50$) $\xi=10^{10}\lambda$, which leads to $r=10^{-13}\lambda^{-1}$. With the Chern--Simons term, $r$ can be much higher. We get $n_s=1-\frac{4}{3N}-\frac{\kappa^{2/3}}{3^{1/3} N^{4/3}}$, which (for $N=50$) gives $n_s=0.973-0.04\kappa^{2/3}$ and $r=10^{-2}\kappa^{2/3}$. The bound $r<0.036$ implies $\kappa^{2/3}<3.6$, and thus $n_s>0.96$. The amplitude of the perturbations $\frac{1}{24\pi^2}\frac{V/F^2}{\epsilon}=2\times10^{-9}$ gives $\xi=3\times10^4 \sqrt{\lambda} \kappa^{-1/3}>1.5\times10^4\sqrt{\lambda}$. This is much smaller than in the case without $P$, and of the same order of magnitude as in the metric Higgs inflation case.

If we instead take the potential and the non-minimal coupling to also go up to dimension 6 (like the Chern--Simons contribution) and assume that the largest power dominates, \ie $V\propto\varphi^6$, $F\propto\varphi^4$, we have $\tilde K\propto\varphi^{-2}$, so $\chi$ is exponential in $\varphi$, as in the original Higgs inflation case. However, now the potential is not asymptotically flat, but exponentially suppressed, $V\propto e^{-2 A \chi}$, where $A$ is a constant, giving the spectral index $n_s=1-4 A^2$. This agrees with observations for $A\approx0.1$, but to get the right number of e-folds, $\chi$ cannot be very large, so keeping only the leading term of the series expansion is questionable. Also, the tensor-to-scalar ratio $r=32 A^2$ is well in excess of the observational upper limit. If we instead keep terms up to dimension $n$ in the potential, the non-minimal Ricci coupling and the non-minimal Chern--Simons coupling, $V\propto\varphi^n$, $F\propto\varphi^{n-2}$, $P\propto\varphi^{n-4}$, the situation remains qualitatively the same, with $\tilde K\propto\varphi^{-2}$, and the potential is exponentially suppressed as $V\propto e^{-(n-4) A \chi}$. Obtaining a nearly scale-invariant spectral index and the right number of e-folds thus becomes increasingly difficult with increasing $n$. In this case, the Chern--Simons term does not solve the problem that higher order terms spoil the flatness of the potential.

The above results for the tensor-to-scalar ratio are based on the assumption that the gravitational wave amplitude depends on the slow-roll parameters in the same way as in standard slow-roll inflation. However, in our case the dynamics of the tensor modes are modified by the Weyl tensor terms in the action \re{actionfinal}.

\subsection{Tensor and vector perturbations}

\subsubsection{Tensor perturbations} \label{sec:tensor}

Unlike for scalar perturbations, for gravitational waves we cannot straightforwardly apply the usual inflationary results for the observables in terms of the inflationary potential, because the Weyl terms in the action \re{actionfinal} change the relation between the background evolution and the tensor perturbations. The parity-violating term that is linear in $P$ has been well studied in the metric case, and it leads to birefringence and instability for one polarisation mode in the deep sub-Hubble regime \cite{Lue:1998mq, Choi:1999zy, Alexander:2004us, Alexander:2004xd, Alexander:2004wk, Lyth:2005jf, Alexander:2006lty, Alexander:2009tp, Dyda:2012rj, Alexander:2016hxk, Bartolo:2017szm, Bartolo:2018elp, Sulantay:2022sag, Creque-Sarbinowski:2023wmb}. In the Palatini case, the terms quadratic in $P'$, which do not break parity, turn out to make the deep sub-Hubble regime stable, although the analysis is limited by the validity of the gradient approximation. Let us see how this comes about. The metric with tensor perturbations is
\bea
  \text{d}s^2 = -\text{d}t^2 + a(t)^2 \left( \delta_{ij} + 2 h_{ij} \right) \rmd x^i \rmd x^j \ ,
\eea
where $h^i{}_i=0$, $\partial_i h^{ij} = 0$, and spatial indices are raised and lowered with the Euclidean metric $\d_{ij}$. The second order tensor part of the action \re{actionfinal} reads
\bea \label{actionTensors}
\nonumber
  S_{\text{tensor}} &=& \int \rmd^4 x a^3 \bigg\{ - \frac12 \bar g^{\a\b} \pat_\a h_{ij} \pat_\b h^{ij} + 2 \dot P \epsilon^{ijk} a^{-1} \bar g^{\a\b} \pat_\a h_i{}^l \pat_\b \pat_j h_{kl} \el
   && - \frac12 ( \a_1 + 4 \a_2 ) \dot P^2
   \Big[
    a^{-4}
   \nabla^2 h_{ij}
   \nabla^2 h^{ij}
  + a^{-2} ( 4 \dot h_{ij} \nabla^2 \dot h^{ij}
   + 2 H \dot h_{ij} \nabla^2 h^{ij}
   + 2 \ddot h_{ij} \nabla^2 h^{ij} )
\el
   && + 
   \ddot h_{ij} \ddot h^{ij} 
   + 2 H \dot h_{ij} \ddot h^{ij}
   + H^2 \dot h_{ij} \dot h^{ij}
   \Big] 
   \bigg\} \  ,
\eea
where $\bar g_{\a\b}$ is the background FLRW metric, dot denotes derivative with respect to $t$, $H\equiv\dot a/a$, and $\epsilon^{ijk}$ is the Euclidean Levi--Civita tensor. The Palatini term proportional to $\dot P^2$ involves higher time derivatives of $h_{ij}$ and thus new degrees of freedom, which are outside the range of validity of our effective theory. As in the scalar case, we reduce the order of the equations by using the lower order equations in the action. At zeroth order in $\dot P$, the equation of motion is
\bea
  \ddot h_{ij} = - 3 H \dot h_{ij} + \frac{1}{a^2} \nabla^2 h_{ij} \ .
\eea
Inserting this into \re{actionTensors}, we obtain
\bea \label{redtensor}
  && \!\!\!\!\!\!\!\! S_{\text{tensor}} =
  \nonumber
  \int \rmd^4 x a^3 \bigg\{ - \frac12 \bar g^{\a\b} \pat_\a h_{ij} \pat_\b h^{ij} + 2 \dot P \epsilon^{ijk} a^{-1} \bar g^{\a\b} \pat_\a h_i{}^l \pat_\b \pat_j h_{kl} \el
   && \!\!\!\!\!\!\!\! - 2 ( \a_1 + 4 \a_2 ) \dot P^2
   \Big[ 
    a^{-4} \nabla^2 h_{ij} \nabla^2 h^{ij}
   + a^{-2} ( \dot h_{ij} \nabla^2 \dot h^{ij}
   - 2 H \dot h_{ij} \nabla^2 h^{ij} )  + H^2 \dot h_{ij} \dot h^{ij}
   \Big] 
   \bigg\} \  ,
\eea
The second term on the first line leads to parity-dependent solutions, so it is convenient to decompose $h_{ij}$ into right- and left-handed polarisation modes (we follow the notation of \cite{Alexander:2004wk}),
\bea \label{Fourier}
  h_{ij}(t, \bx) = \frac{1}{(2 \pi)^{3 / 2}} \int \mathrm{d}^3 k \sum_{I=R,L} h_I(t, \bk) p_{ij}^I(\bk) \mathrm{e}^{i \bk\cdot\bx} \ ,
\eea
where $h_I(t, \bk)$ is the amplitude and $p_{ij}^I(\bk)$ is the polarisation tensor, which satisfies $p_{ij}^R(\bk)^*=p_{ij}^L(\bk)$, $p_{ij}^I(\bk) p^{I\, ij}(\bk)=0$, $p_{ij}^R(\bk) p^{L\, ij}(\bk)=2$. Inserting \re{Fourier} into the action \re{redtensor}, we obtain
\bea
  \!\!\!\!\!\!\!\!\! S_{\text{tensor}} &=& \int \rmd t \rmd^3 k a^3 \sum_{I=R,L}  \Bigg\{ \dot h^*_I \dot h_I  \left[ 1 - 4 \lambda_I \dot P \frac{k}{a} - 4 ( \a_1 + 4 \a_2 ) \dot P^2 \left( H^2 - \frac{k^2}{a^2} \right) \right] \el
  && - \frac{k^2}{a^2} h^*_I h_I \left[ 1 - 4 \lambda_I \dot P \frac{k}{a} - 4 ( \a_1 + 4 \a_2 ) \dot P^2 \left( H^2 + \dot H + 2 H \frac{\ddot P}{\dot P} - \frac{k^2}{a^2} \right) \right] \Bigg\} \ ,
\eea
where $\lambda_R = +1, \lambda_L = -1$. The factor $\a_1+4\a_2$ is positive ($4$ in the unconstrained and the zero non-metricity case, $3$ in the zero torsion case). The term proportional to $\lambda_I \dot P$ on the first line is the same as in the metric case: there it makes the kinetic term of one of the polarisations negative for $k/a\gtrsim1/|\dot P|$ \cite{Lyth:2005jf, Alexander:2004wk, Alexander:2009tp, Dyda:2012rj, Bartolo:2017szm, Bartolo:2018elp, Creque-Sarbinowski:2023wmb}. In the Palatini case the terms proportional to $\dot P^2$ cure the instability in the regime $k/a\gg|\dot P|$. However, there is an instability for both polarisations for momenta $k/a\lesssim H$ unless $|\dot P| H\lesssim0.2$. Demanding $|\dot P| H\ll1$ puts the theory far from the instability for all momenta. Then the evolution of super-Hubble modes is standard, and the Bunch--Davies vacuum mode remains a solution deep in the sub-Hubble regime, but propagation from the sub-Hubble to the super-Hubble regime is strongly modified. This could change the predictions for gravitational waves. However, because we have used the gradient approximation, it is questionable whether the effective theory can be trusted when $\dot P^2 k^2/a^2\gg1$. Within the regime where the theory is valid, the corrections are small.

\subsubsection{Vector perturbations}

Let us now consider vector perturbations, described by the metric 
\bea \label{vectormetric}
  \text{d}s^2 = - \text{d}t^2 + 2 a(t) B_i \rmd t \rmd x^i + a(t)^2 \delta_{ij} \rmd x^i \rmd x^j \ ,
\eea
with $\pat_i B^i=0$; spatial indices are again raised and lowered with the Euclidean metric $\delta_{ij}$. It has been claimed that vector perturbations do not contribute to the metric part of the Chern--Simons term \cite{Alexander:2004us}. This is not correct, but in the metric case the Chern--Simons term does not introduce time derivatives of $B_i$, so vector perturbations remain non-dynamical and are not sourced. With the Palatini contribution to the Chern--Simons term, the situation is different. Inserting the metric \re{vectormetric} into the action \re{actionfinal}, using the background equations, and keeping only the second order vector part, we have
\bea \label{actionvectors}
    S_{\textrm{vectors}}
   &=&
   \int \rmd^4 x a^3
   \bigg[ - \frac{1}{4}
   \frac{1}{a^2}
   B_i \nabla^2 B^i
   + \frac12 \frac{1}{a^3} \dot P
   \epsilon^{ijk}
   B_i
   \partial_j \nabla^2
   B_k
   \el
&& - \frac14 ( \a_1 + 4 \a_2 ) \dot P^2 
   \left( \frac{1}{a^2}
   \dot B_i
   \nabla^2 
   \dot B^i
   + \frac{1}{a^4}
   \nabla^2 B_i
   \nabla^2 B^i
   \right) 
   \bigg] \ .
\eea
The Palatini contribution on the second line, taken at face value, leads to a dynamical vector field (when $\dot P\neq0$). However, the order of a differential equation changing when a perturbative parameter goes to zero signals the presence of unphysical solutions that should be excised by the order reduction procedure \cite{Bel:1985zz, Simon:1990ic, Simon:1990jn, Simon:1991bm, Parker:1993dk, Weinberg:2008hq, Delsate:2014hba, Solomon:2017nlh}. In the gravitational wave case, we found that the validity of the theory is restricted to momenta $k/a\ll1/|\dot P|$. Applying the same condition here would remove the metric contribution to the Chern--Simons term as well as the second Palatini contribution term on the second line. However, the first term on the second line, which makes the vector field dynamical, is not suppressed by this factor, as it has the same number of spatial derivatives as the leading term that comes from the Einstein--Hilbert part of the action. Applying the order reduction technique gets rid of all the Chern--Simons contributions, as the leading order solution for $B_i$, and hence the whole reduced action, vanishes.

\section{Conclusions} \label{sec:conc}

We have for the first time studied the Chern--Simons term coupled to the inflaton $\varphi$ via the function $P(\varphi)$ in the Palatini formulation. In contrast to the metric formulation, the Chern--Simons term modifies the background and scalar perturbations, not only gravitational waves. As in the metric case, the theory is likely unstable, and has to be considered as an effective theory. We use the gradient approximation, and reduce the order of the equations of motion by substituting the leading order equations back into the action. We consider three cases: when the connection is unconstrained, and when either non-metricity or torsion is taken to vanish a priori. To leading order, in the zero torsion case there is no change in the inflationary background and scalar perturbations. In contrast, in the unconstrained and the zero non-metricity case the inflaton kinetic term gets an additive modification proportional to $P'{}^2$ times the square of the inflaton potential. This is reminiscent of New Higgs Inflation, where a term proportional to the potential is added to the kinetic term \cite{Germani:2010gm, DiVita:2015bha, Escriva:2016cwl, Fumagalli:2017cdo, Gialamas:2020vto, Nezhad:2023dys}, but here there is no direct kinetic coupling in the original action, the modification is generated through the order reduction procedure.

We consider the effect of this modification on some example inflationary models. For the potential $V\propto\varphi^n$ (with $n\geq2$) and coupling function $P$ that is either quadratic or chosen so that the dimension of the Chern--Simons term is $n$ (for $n\geq6$), we get the potential $V\propto\chi^\a$ for the canonically normalised field, with $1>\a\geq\ha$. This is similar to the effect of non-minimal couplings in Palatini gauge field theory \cite{Rasanen:2022ijc}. This solves the problem of higher order corrections spoiling the flatness of the inflationary potential: as the original potential becomes steeper, the kinetic term adjusts correspondingly to keep the potential for the canonical field flat. While the resulting spectral index agrees with observations, the tensor-to-scalar ratio $r$ remains too high, except possibly for $\a\approx\ha$, when it would be on the verge of discovery. However, this is subject to uncertainties due to preheating, and the result could be different if more than one power contributes to the potential and the coupling function. We also consider Higgs inflation. In the standard case without the coupling to the Chern--Simons term, the potential for $\chi$ is exponentially asymptotically flat, with an unobservably small $r$. With the simplest coupling function $P\propto\varphi^2$, the potential approaches flatness only as $1/\chi$, so $r$ can take any value. The non-minimal coupling to the Ricci scalar $\xi$ can be as small as in the metric case, $\xi\sim10^4$, compared to the usual Palatini value $\xi\sim10^8$.

At leading order in the gradient approximation, perturbations are related to the background evolution in the same way as in the case without the Chern--Simons term. For vectors and scalars, corrections to this appear at a lower order in the gradient approximation than for scalars due to a term linear in $P'$ that breaks parity and is present also in the metric case. There are also parity-conserving terms quadratic in $P'$ that appear only in the Palatini case. They cure the instability of one polarisation of the tensor modes that exists in the metric case. In the Palatini case, both polarisations are stable in the sub-Hubble regime, but there can be an instability for modes with wavelengths close to or larger than the Hubble scale, depending on the amplitude of $P'$. If this instability is not present, the tensor modes are stable for all momenta, with large modifications to sub-Hubble propagation. However, this happens only when the gradients are large and the validity of our approximation is questionable. The issue requires more detailed study. As for vector perturbations, the Palatini contribution to the Chern--Simons term would naively make them dynamical, but such an effect is beyond the scope of our effective theory.

For three- and four-point functions and higher order statistics, in the Palatini formulation there are new terms to consider beyond those present in the metric Chern--Simons case \cite{Bartolo:2017szm, Bartolo:2018elp, Creque-Sarbinowski:2023wmb}, so observations of non-Gaussianity may provide a distinctive signature for the Chern--Simons coupling in the Palatini formulation. Given the recently claimed detection of parity violation in the four-point function of large-scale structure \cite{Hou:2022wfj, Philcox:2022hkh} (see also \cite{Krolewski:2024paz}), it would be interesting to calculate this signal. Confirming the instability of the theory also remains an open question. 

\section*{Acknowledgments}

AH acknowledges funding from the Finnish National Agency for Education (EDUFI).

\bibliographystyle{JHEP}
\bibliography{cs}

\end{document}